\documentclass{article}
\usepackage{amsmath,amssymb}
\usepackage{graphicx}%
\usepackage{verbatim}
\begin{document}

\begin{center}
{\bf \Large Averaging of the Equations of the Standard Cosmological Model over Rapid Oscillations:\\ Influence of the cosmological term on the mean value of the effective barotropic coefficient}. \\[12pt]
Yurii Ignat'ev$^1$ and A. R. Samigullina$^2$ \\
$^1$ Institute of Physics, Kazan Federal University; $^2$ N.I. Lobachevsky Institute of Mathematics and Mechanics, Kazan Federal University, \\ Kremleovskaya str., 35, Kazan, 420008, Russia\footnote{Yurii Ignat'ev and A. R. Samigullina, Russ.Phys.J., {\bf 61}, 643 (2018).}
\end{center}

\begin{abstract}
With the help of an applied software package written by the authors, we have averaged the effective total barotropic coefficient $\kappa=(-\lambda + p)/(\lambda+\epsilon$) of the classical scalar field and the cosmological term and have shown that during cosmological evolution for sufficiently large values of the cosmological constant the Universe transitions from the inflationary stage to a nonrelativistic stage, and then, after a plateau, it transitions to a later inflationary stage.

{\bf keywords}{\it standard cosmological model, averaging of invariant characteristics, mean value of the cosmological acceleration, stage of cosmological expansion, numerical gravitation.}

\end{abstract}

\section{Introduction}
In previous works \cite{Ig1,Ig2} it was shown with the help of averaging of rapidly oscillating numerical solutions of the equations of cosmological evolution over a macroscopic time interval for the standard cosmological model with zero cosmological term that over the course of cosmological time the contribution to the energy density from microscopic quadratic oscillations of the scalar field begins to dominate over the contribution from the averaged macroscopic scalar field, and that the ratio of these two contributions reaches values on the order of   $10^4$ at later times.  Simultaneously with this, the effective macroscopic equation of state tends to its nonrelativistic limit, i.e., the mean value of the barotropic coefficient $\kappa =p/\varepsilon $ tends to zero.  This phenomenon was interpreted in \cite{Ig1,Ig2} as a process of production of nonrelativistic scalar bosons.  In the case of a nonzero value of the cosmological constant $\lambda $ the invariant cosmological acceleration $\Omega $ is determined not only by the barotropic coefficient
\begin{equation} \label{Eq1}
\Omega =-\frac{1}{2} (1+3\kappa ),
\end{equation}
but also depends on the magnitude of the cosmological constant.  Taking the cosmological constant into account, the total barotropic coefficient should be redefined for the classical scalar field
$\Phi (\tau )$ as follows\footnote{The Planck system of units is used: $G=c=\hbar =1$, and for the time variable, the dimensionless variable $ \tau =mt$  is used, where m  is the mass of the scalar field (for details, see\ \cite{1}.}:
\begin{equation} \label{Eq2}
\kappa =\frac{-\lambda +p}{\lambda +\varepsilon} =\frac{-\lambda +\Phi'^{2} -\Phi ^{2}}{\lambda +\Phi '^{2} +\Phi ^{2}} ,
\end{equation}
where $f'\equiv df/d\tau$. In light of this, the need arises to investigate the influence of the cosmological constant on the evolution of the mean value of the total barotropic coefficient of the standard cosmological model. Numerical modeling was performed in this work on the basis of the authors' updated software package DifEqTools \cite{Ig3} free access to which is open on the website [4]\footnote{This work was performed according to the Russian Government Program of Competitive Growth of Kazan Federal University.}.

\section{Main relations of the mathematical model}
The full system of equations of the standard cosmological model for the spatially flat Friedmann model (the model considered here) in the chosen system of units has the following form:
\begin{equation} \label{Eq3}
3\frac{a'^{2} }{a^{2} } =\lambda _{m} +\Phi '^{2} +\Phi ^{2} \Rightarrow H_{m} {}^{2} \equiv \Lambda '^{2} =\frac{1}{3} (\lambda _{m} +\Phi '^{2} +\Phi ^{2} )
\end{equation}
is the only nontrivial Einstein equation, where
\begin{equation} \label{Eq4}
\Lambda =\ln (a);\frac{a'}{a} \equiv \Lambda '=H{}_{m} (t)
\end{equation}
$(H(t)$ is the Hubble constant), and the equation of the classical massive scalar field $(\dot{f}\equiv df/dt)$ is
\begin{equation} \label{Eq5}
\Phi ''+3H_{m} \Phi '+\Phi =0.
\end{equation}
Here the energy-momentum tensor (\ref{Eq1}) has the structure of the energy-momentum tensor of an isotropic fluid with \textit{reduced} energy density and pressure:
\begin{equation} \label{Eq6}
\varepsilon =\frac{1}{8\pi} \left(\Phi '^{2} +\Phi ^{2} \right); p=\frac{1}{8\pi} \left(\Phi '^{2} -\Phi ^{2} \right),
\end{equation}
so that
$$\varepsilon +p=\frac{\Phi '^{2} }{4\pi } .$$

Expressing the Hubble constant $H_{m}^{} (t)$ from the Einstein equation (\ref{Eq3}) in terms of the functions $\Phi$ and $\Phi'$ and carrying out the standard substitution of variables $\Phi'=Z(\tau )$, we reduce system of field equations \ref{Eq3} and \ref{Eq6} to the form of a normal autonomous system of ordinary differential equations in the three-dimensional phase space $\{\Lambda ,\Phi ,Z\} $:
\begin{equation} \label{Eq7}
\Lambda '\equiv h=\sqrt{\frac{1}{3} \left(\lambda _{m} +Z^{2} +\Phi ^{2} \right)} ,
\end{equation}
\begin{equation} \label{Eq8}
\begin{array}{l}
{\Phi '=Z} \end{array},
\end{equation}
\begin{equation} \label{Eq9}
Z'=-\sqrt{3} \sqrt{\lambda _{m} +Z^{2} +\Phi ^{2} } Z-\Phi .
\end{equation}
Here
\begin{equation} \label{Eq10}
H=m\frac{a'}{a} \equiv mh,\Omega =\frac{aa''}{a'^{2} } \equiv 1+\frac{h'}{h^{2}}.
\end{equation}
System of autonomous ordinary differential equations \ref{Eq7}--\ref{Eq9} has an autonomous subsystem in the $(\Phi ,Z)$ plane, which we shall investigate.\\

\section{Calculation of time-averaged functions of the dynamic variables}

We shall calculate the mean values of the functions of the dynamic variables according to the rule
\begin{equation} \label{Eq11}
\overline{\psi }(\tau )=\frac{1}{\Delta \tau } \int _{\tau }^{\tau +\Delta \tau }\psi (\tau ')d\tau ' ,
\end{equation}
where $1\ll \Delta \tau \ll \tau $ is the averaging interval.  What is of particular interest to us in this work is the mean value of the barotropic coefficient (\ref{Eq2}):
\begin{equation} \label{Eq12}
\bar{\kappa }=\overline{\left(\frac{-\lambda +\Phi '^{2} -\Phi ^{2} }{\lambda +\Phi '^{2} +\Phi ^{2} } \right)}.
\end{equation}
The above-indicated authors' software package DifEqTools  \cite{Ig4} allows one to calculate averages of functions of dynamic variables automatically directly with numerical solution of the system of nonlinear differential equations using the command

 $\overline{\varphi }$(s) = DifEqTools[NumericDsolveMiddle](EQS, ICS, Methods, [$\tau $,$\varphi (\tau )$], $\Delta \tau $, s, N, M),

where EQS is the list of differential equations, ICS is the list of initial conditions, Methods is the list of methods and parameters of the numerical integration procedure, [$\tau $,$\varphi (\tau )$] is a list consisting of an independent variable and the function to be averaged, $\Delta \tau$ is the duration of the averaging interval, N is the number of bins the interval is partitioned into, and  M is the integration method. We present here an example of the use of this command, corresponding to the performed calculations:

\verb|Kappa_Mid:=(t,Lambda,Phi0) -$\mathrm\verb|.
\begin{verbatim}
>DifEqTools[NumericDsolveMiddle](EqF(Lambda),ICF(Phi0),
 [method=dverk78,maxfun=500000,abserr = 1.*10^(-6),
 relerr = 1.*10^(-6)],
 [tau, kappa(tau,Lambda)], 4*Pi,t,100,S);
\end{verbatim}
Figures 1 -- 3 show some numerical modeling results obtained using this software package.
\begin{center}
\includegraphics[width=12cm]{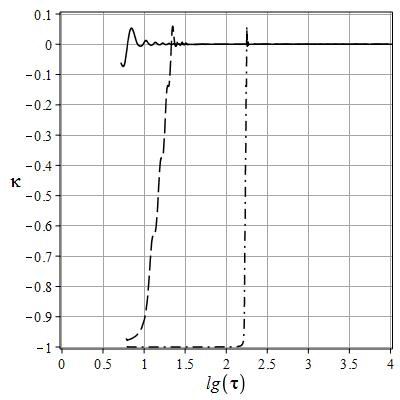}\\ \noindent {\bf Fig.\refstepcounter{figure} \thefigure.}\quad
Influence of the initial value of the scalar potential on the evolution of the mean value of the effective barotropic coefficient: \textit{a}) for $\lambda _{m} =0$, the solid curve corresponds to $\Phi _{0} =1$, the dashed curve corresponds to $\Phi _{0} =10$, and the dot-dash curve corresponds to $\Phi _{0} =100$, \textit{b}) for small values of $\lambda _{m} =10^{-12} $, from left to right: $\Phi _{0} =1$, $\Phi _{0} =10$, $\Phi _{0} =100$.
\end{center}

\begin{center}
\includegraphics[width=12cm]{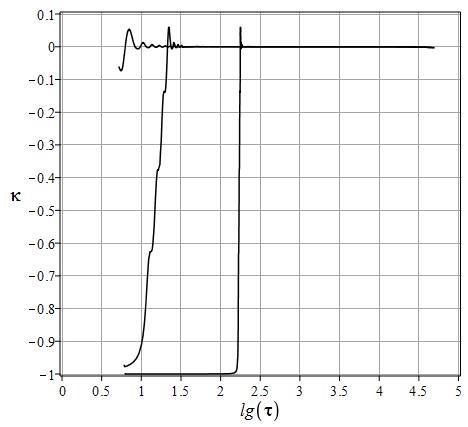}\\ \noindent {\bf Fig. \refstepcounter{figure}\thefigure.}\quad
Effect of the initial value of the scalar potential on the evolution of the average effective barotropic coefficient for small values $\lambda_m=10^{-12}$. From left to right: $\Phi _{0} =1$, $\Phi _{0} =10$, $\Phi _{0} =100$.
\end{center}

\begin{center}
\includegraphics[width=12cm]{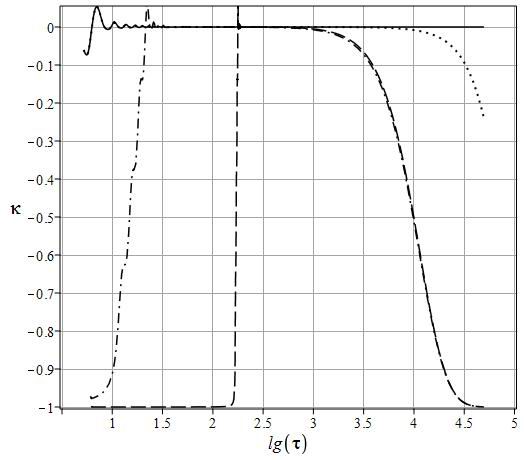}\\ \noindent {\bf Fig. \refstepcounter{figure}\thefigure.}\quad
Influence of the magnitude of the cosmological constant on the evolution of the mean value of the effective barotropic coefficient: the solid curve corresponds to $\lambda =0,\Phi _{0} =1$, the dotted curve corresponds to $\lambda =10^{-10},\Phi _{0} =1$, the long-dash curve corresponds to $\lambda = 0.001,\Phi _{0} =1$, the dot-dash curve corresponds to $\lambda =10^{-8},\Phi _{0} =10$, and the dashed curve corresponds to $\lambda =10^{-8},\Phi _{0} =100$. At late stages the two last curves coalesce.
\end{center}

Thus we see that for very small values of the cosmological constant the mean value behaves similarly to the model with zero value of the cosmological constant: in the early expansion stage $\kappa \to -1$, then the mean value of the barotropic coefficient enters into an oscillatory regime near its zero value.  Corresponding to this, the invariant cosmological acceleration falls from +1 to -1/2, i.e., cosmological matter evolves from the inflationary state to the nonrelativistic state. With increase of the value of the cosmological constant, the behavior of the cosmological model \textit{on a macroscopic time scale} changes substantially: the mean value of the barotropic coefficient varies from -1 to 0, and then, after a plateau, falls back to -1. Thus, in macroscopic time the Universe transitions from early inflation to a nonrelativistic stage, and from there to late inflation.

\end{document}